\documentstyle[12pt,epsfig]{article}
\begin{document}
\pagestyle{empty}
\renewcommand{\thefootnote}{\fnsymbol{footnote}}
\def\lsim{\raise0.3ex\hbox{$<$\kern-0.75em\raise-1.1ex\hbox{$\sim$}}}
\def\gsim{\raise0.3ex\hbox{$>$\kern-0.75em\raise-1.1ex\hbox{$\sim$}}}
\def\propsim{\raise0.3ex\hbox{$\propto$\kern-0.75em\raise-1.1ex\hbox{$\sim$}}}
\def\noi{\noindent}
\def\nn{\nonumber}
\def\bea{\begin{eqnarray}}  \def\eea{\end{eqnarray}}
\def\beq{\begin{equation}}   \def\eeq{\end{equation}}
\def\beeq{\begin{eqnarray}} \def\eeeq{\end{eqnarray}}
\def\R{ {\rm R \kern -.31cm I \kern .15cm}}
\def\C{ {\rm C \kern -.15cm \vrule width.5pt \kern .12cm}}
\def\Z{ {\rm Z \kern -.27cm \angle \kern .02cm}}
\def\N{ {\rm N \kern -.26cm \vrule width.4pt \kern .10cm}}
\def\1{{\rm 1\mskip-4.5mu l} }
\def\lsim{\raise0.3ex\hbox{$<$\kern-0.75em\raise-1.1ex\hbox{$\sim$}}}
\def\gsim{\raise0.3ex\hbox{$>$\kern-0.75em\raise-1.1ex\hbox{$\sim$}}}
\def\sq{\hbox {\rlap{$\sqcap$}$\sqcup$}}
\vbox to 2 truecm {}
\centerline{\Large \bf $x_F$-dependence of $J/\Psi$ suppression in
$pA$ collisions.}

\vskip 1 truecm
\centerline{\bf C. A. Salgado\footnote{e-mail: carlos.salgado@th.u-psud.fr}}
\centerline{Laboratoire de 
Physique Th\'eorique\footnote{Unit\'e Mixte de Recherche -
CNRS - UMR n$^{\circ}$ 8627}}  
\centerline{Universit\'e de Paris XI, B\^atiment 210,
F-91405 Orsay Cedex, France}

\begin{abstract}
Coherence effects are important in the description of collisions with
extended objects as nuclei. At fixed target energies and small $x_F$, the
coherence length of the fluctuation containing the $c\bar c$ is
small and the usual nuclear absorption model is valid. 
However, at
higher energies and/or
$x_F$ the nucleus is seen as a whole by the fluctuation.
In this case, the total, not the
absorptive, $c\bar c-N$ cross section controls the suppression and also
shadowing of gluons appears. 
We propose that the growth of the coherence length 
can explain the $x_F$-dependence of present 
experimental data. For this, we need a ratio of absorptive over 
total $c\bar c-N$ cross section of 0.2.


\end{abstract}

\vskip 1 truecm

\noindent LPT Orsay 01-41\par
\noi \today \par

\newpage
\pagestyle{plain}
\baselineskip=24 pt

The $J/\Psi$ suppression is one of the main signals for quark gluon plasma (QGP)
formation \cite{matsuisatz}. 
The {\it anomalous suppression} observed by NA50 Collaboration
\cite{na50} is interpreted as produced by a deconfined state \cite{deconf}
though hadronic interpretations are possible \cite{alfons}. Whether NA50 data
give or not a definite proof of the formation of 
a deconfined state is
a topic of intense discussion. In this letter we study
the {\it non-anomalous suppression} in the whole range of $x_F$
a subject that, though at first sight
may seem solved, has some remaining open problems. This {\it normal
suppression} is usually ascribed to multiple interaction of the
produced pre-resonant $c\bar c$ state (color octet \cite{kharzeev}) 
with the surrounding 
nuclear matter. The picture is very simple and well known: a $c\bar c$
is created at some point $z_0$ inside a nucleus in an octet state.
In its travel through the nucleus, this state can interact at
points $z>z_0$ with other nucleons that will destroy it with a 
cross section $\sigma^{abs}$. The formula describing this nuclear
absorption is, after integrating in $z$:

\beq
\sigma_{pA} = {\sigma_{pp}\over\sigma^{abs}} \int d^2b \left
[ 1 - \exp \left ( - \sigma^{abs} A T_A(b) \right ) \right ] \ \ \ .
\label{eqnucabs}
\eeq

\noi where $T_A(b)=\int dz  \rho_A(b,z)$ is the profile function for 
nucleus A normalized to 1 and $\rho_A(b,z)$ is the nuclear density,
that we take from ref.
\cite{densities}.

This formula describes well the observed $pA$ data at midrapidities 
measured by NA38-NA51 \cite{pAdata} both for $J/\Psi$ and
$\Psi'$ suppression, supporting the interpretation of the preresonant
color octet state. A cross section of $\sigma^{abs}=6.5\pm 1.0$ mb is 
obtained by the experimental collaboration. 
This cross section does not depend on the energy.
For this analysis, data from
$OCu$, $OU$ and $SU$ collisions were also included. 
The 
E866/NuSea Collaboration \cite{e866} has also measured $J/\Psi$ suppression
in $pA$ collisions and the result does not fully agree with the one from
NA38-NA51:
using the parametrization $\sigma^{pA}=A^\alpha
\sigma^{pp}$ one obtains $\alpha\sim 0.95$ for E866/NuSea \cite{e866} at 
$y\sim0$
and $\alpha\sim 0.92$ for CERN data. 
To reproduce E866 data at $x_F\sim 0$ a $\sigma^{abs}\sim$ 3 mb is needed.
E866/NuSea data give also the 
$x_F$ dependence of the nuclear suppression, the origin of which can not be
completely attributed 
to modifications of the nuclear gluon distributions \cite{eks98} 
(see bellow) and
remains an open problem. 
The main goal of the present work is to describe this $x_F$ dependence
and also to compare the different experiments. Understanding the rapidity
pattern of the absorption is very important in order to have
a real knowledge of the physics behind the suppression and also to 
extrapolate to RHIC and LHC energies, where this nuclear absorption
would be present.

The idea is simple and its theoretical formulation has been derived 
in a previous paper \cite{mij}.
In the frame where the nucleus is at rest, the incoming proton fluctuates
in a complicated system of quarks and gluons with coherence length $l_c$.
At small energies and $x_F$, 
$l_c$ is small (of the order of the nucleon size) and
only one nucleon in
the nucleus takes part in the hard interaction that produces the $c\bar c$. 
This implies $\sigma_{pA} 
\sim A$. This behavior is modified by the collisions of the produced
$c\bar c$ with the other nucleons in the nucleus. We have, in this way, the
usual
description of nuclear absorption given by eq. (\ref{eqnucabs}).
However, at large energies and/or $x_F$, 
$l_c$ gets eventually larger than the nuclear size and the nucleus is
seen as a whole by the fluctuation. As a consequence 
the time ordering is lost and (\ref{eqnucabs}) is no longer valid. To describe 
this regime, we have introduced two types of collisions
with the nucleons in the target, the ones of
the light partons (mainly gluons) -- with total cross section $\sigma$ --
and the ones of the heavy system ($c\bar c$) -- with total cross section 
$\widetilde\sigma$ --
in an eikonal approach.
The first ones give rise to modifications in the nuclear gluon distribution
and the second to a suppression of charmonia states (the generalization
of nuclear absorption). The result which replaces eq. (\ref{eqnucabs}) is
\cite{mij}

\begin{equation}
{d\sigma_{pA}\over dx_Fd^2b}
= \sigma_{pQCD}^{gg\to c\bar cX} 
g_{p}(x_{1},Q^{2}) g_{A}(x_{2},Q^{2}, b)
e^{-{1 \over 2} \widetilde{\sigma} AT_A(b)},
\label{eqNPB3}
\end{equation}

\noi
with $x_F=x_1-x_2$, $x_1x_2 s=m_{J/\Psi}^2$, and

\begin{equation}
g_A(x_2, Q^2, b)
= 2 \int d\omega \left [ 1 - e^{-{1 \over 2}
\sigma(\omega) A T_A(b)} \right ] .
\label{eqNPB4}
\end{equation}

\noi $\omega$ represents kinematical variables of the gluon-nucleon
interactions to be integrated. This term gives the modification of nuclear 
structure functions with respect to nucleons and will not be discussed here --
see for instance \cite{perc} for a model. In (\ref{eqNPB3}) we see that
the multiple scatterings of the heavy and light systems factorize, so that
we can separate both contributions in the ratios $R_{pA}$ of 
$pA$ to $pp$ cross sections.

\beq
R_{pA}=R_{pA}^{shadow} R_{pA}^{c\bar c}.
\label{eqratios}
\eeq

Neglecting shadowing corrections to gluons, 
the change from low to asymptotic energies consists in the substitution:

\begin{equation}
{1\over\sigma^{abs}} \left [ 1 - \exp \left ( -\sigma^{abs} \
A
 \ T_A(b)
\right ) \right ] \longrightarrow
A\ T_A(b) \exp \left ( - {1 \over 2} \widetilde{\sigma} \
 A \
T_A(b) \right ),
\label{eqNPB2}
\end{equation}

The most important point is the change of the absorptive cross section
by the total one. When $\widetilde{\sigma}=\sigma^{abs}$ the first correction 
term in the expansion
in $\widetilde{\sigma}$ is the same for both expressions.
As these cross sections are not very large in practice, 
the numerical values turn out to be
very similar.
In this case, this justifies the use of formula (\ref{eqnucabs}) for
high energy though strictly speaking it is only valid at small energies.
In our case, however, 
we will suppose that $\widetilde{\sigma}\neq \sigma^{abs}$. That is,
the preresonant $c\bar c$ state has a non negligible probability
of not being destroyed.
This increases the absorption for large values of $x_F$ where coherence
is reached and $\widetilde{\sigma}$ has to be used. 

The two regimes are particular solutions of a more general equation
which takes into account the coherence effects for any $l_c$.  
In this case the factorization given in (\ref{eqratios}) is no longer valid.
Neglecting shadowing to structure functions,

\begin{equation}
{d\sigma_{A}\over dx_F}
 = \sigma_{pQCD}^{gg\to c\bar cX}
g_{p}(x_{1},Q^{2}) g_{p}(x_{2},Q^{2})
\ \sum_{n=1}^A C_A^n
\sum_{j=1}^n
\int d^2 bT_n^{(j)} (b) \sigma_n^{(j)} \ \ \ ,
\label{eqNPB28}
\end{equation}

\noindent
where
\begin{equation}
\sigma_n^{(j)} =  
j\left (-{\widetilde{\sigma}\over 2}\right)^{j-1}
(-\sigma^{abs})^{n-j}
+ (j-1)\left (-{\widetilde{\sigma}\over 2}\right)^{j-2}
(-\sigma^{abs})^{n-j+1}\ .
\label{eqNPB30}
\end{equation}

\noi 
The powers of nuclear profile functions in the expansion of (\ref{eqNPB3})
have to be changed to
\begin{equation}
T^n(b)\ \ \  \longrightarrow \ \ \ 
T_{n}^{(j)}(b)=
n!\int_{-\infty}^{+\infty}dz_{1}\int_{z_{1}}^{+\infty}dz_{2}...
\int_{z_{n-1}}^{+\infty}dz_{n}\cos(\Delta
(z_{1}-z_{j}))\prod_{i=1}^{n} \rho_{A}(b,z_{i}).
\label{eqNPB25}
\end{equation}

\noi The physical interpretation of the above equations is very clear,
the first $j-1$ collisions are coherent (one of them
corresponding to the light amplitude $\sigma$), the last $n-j$ are not
and the $j$-th has the two possibilities.
The $\Delta$ factor that controls these finite energy effects can be 
approximated by

\begin{equation}
\Delta \equiv {1\over l_c}=  {m_p M^2\over s \ x_1} \ \ \ ,
\label{eqNPB26}
\end{equation}

\noi where $m_p$ is the mass of the proton and
$M^2$ is the effective mass of the fluctuation. This effective mass 
could be measured in diffractive events containing a $J/\Psi$. 
In order to obtain the observed scaling of the suppression in $x_F$,
we take $M^2/s\sim $ constant. This means that increasing the energy allows the 
fluctuation to have bigger masses. This is the kind of behavior 
of triple pomeron contribution to diffraction \cite{lowx},
however it is not clear that these are the relevant diagrams in the present
case. We take this scaling as a phenomenological ansatz, having in 
mind that this is most probably a finite energy effect that
would disappear at asymptotic energies\footnote{I thank
A. Capella and A. Kaidalov for discussions on this point.}.
In practice, what we have done is taking $M^2=m_{J/\Psi}^2$ for the smallest 
experimental energy ($\sqrt{s}\sim 20$ GeV) and use the proportionality with
$s$ for the others. 

In our calculations we have also introduced the nuclear corrections to 
parton distributions (gluons) as given by the EKS98 parametrization 
\cite{eks98}. We 
could also use eq. (\ref{eqNPB4}), however this would need some theoretical 
inputs from gluon-nucleon cross section that complicates the computation.
Moreover, gluon antishadowing can not be reproduced by this formula. To 
take into account coherence effects we have defined

\beq
R_{pA}^{shadow}=\left [1-\left (1-{A_{eff}^{fin}-A_{eff}^{asym}\over
A_{eff}^{prob}-A_{eff}^{asym}}\right )\left(1-R_{pA}^{EKS}\right )\right ]
\label{eqshad}
\eeq

\noi $R_{pA}^{EKS}$ being the shadowing corrections to $c\bar c$ production
computed 
with EKS98 parametrizations, and $A_{eff}^{prob}$, $A_{eff}^{asym}$
and $A_{eff}^{fin}$ the effective A (i.e. $\sigma_{pA}/\sigma_{pp}$)
computed with eq. (\ref{eqnucabs}), (\ref{eqNPB3}) and (\ref{eqNPB28})
respectively, all without shadowing. 
In this way, $R_{pA}^{shadow}= 1$ when
$l_c$ is very small ($A_{eff}^{fin}=A_{eff}^{prob}$)
and $R_{pA}^{shadow}=R_{pA}^{EKS}$ for large coherence
lengths ($A_{eff}^{fin}=A_{eff}^{asym}$). 
However, as the data from E866/NuSea uses $Be$ as a reference, the 
shadowing corrections are not large and, in any case, much smaller than the
observed effect (see Fig. 1).
The final result for absorption is given by eq. (\ref{eqratios}) with 
$R_{pA}^{c\bar c}=A_{eff}^{fin}/A$.

With all these ingredients we have fitted the data from E866/NuSea 
Collaboration.
The free parameters are $\sigma^{abs}$ and $\widetilde{\sigma}$. 
To have a good description of E866/NuSea data we obtained $\sigma^{abs}$= 3 mb
and $\widetilde{\sigma}$= 15 mb. The comparison with experimental data
can be seen in Fig. 1\footnote{Actually, only $W/Be$ data have been used in the
fit, the $Fe/Be$ data came out as a result.}. Also shown in this
figure is the comparison taking into account only shadowing corrections
to gluons given by EKS98 parametrization
and nuclear suppression given by eq. (\ref{eqnucabs})
with $\sigma^{abs}$=4.5 mb. Notice that if shadowing (antishadowing 
at $x_F\sim 0$) is not included, $\sigma^{abs}\sim$ 3 mb is needed
in order to reproduce the data at $x_F\lsim 0.2$.

In Fig. 2 we present the comparison with NA38-NA51 data. These data are 
measured in the interval $3<y_{lab}<4$. We have used the values $<x_F>=$ 0.03
for $E_{lab}=$450 GeV data and $<x_F>=$ 0.16 for $E_{lab}=$200 GeV data.
The description of the data is good in spite of the apparent discrepancy
in the parameter $\alpha$ among the two sets discused above. Notice that
the data at 200 GeV is more suppressed due to the larger $x_F$.

Finally, in Fig. 3 comparison is made with NA3 data \cite{na3}
at $E_{lab}=$200 GeV. The
description is again not bad, though some discrepancy appears at
large values of $x_F$. This is also observed in other analysis \cite{arleo}
and could be a signal of energy loss \cite{eloss}, however, the
evidence is too weak and more experimental data would be needed in this 
region\footnote{Notice that energy loss with nuclear matter becomes smaller
with increasing energy \cite{domin}, so the effect in E866/NuSea will be
smaller.}. 
Besides, including shadowing corrections (in 
fact antishadowing in this region) makes our results to increase,
breaking the scaling on $x_F$ mentioned above. This corrections are very
uncertain and this could also be the origin of the discrepancy.

Let's compare our analysis with previous ones. In \cite{arleo} 
a description of data very similar to ours was obtained by taking into 
account two different cross sections for octet and singlet $c\bar c$ states,
the drawback of this model was the smallness of the
lifetime of the octet state. In \cite{hwa}, some kind of energy loss of 
gluons from the incoming proton is proposed. 
In \cite{ramona} different effects as shadowing, energy loss, comover
absorption and intrinsic charm are taken into account separately.
In \cite{kaidalov} a model similar to ours, that takes into account
coherence effects, has been developed. They also
include two types of interactions for light and heavy systems, obtaining
shadowing and $c\bar c$ suppression,
however,
their $x_F$ dependence come from loosing of momentum of the
$c\bar c$ pair as it experiences multiple scattering.
Finally, in \cite{kopnew} coherence effects as well as other effects as
energy loss and time formation
are taken into account. This is a more formal analysis, but where
coherence give effects similar to ours.
Their computations are only for the $\chi_c$ state.
One main difference is that they don't
take into account an effective mass of the fluctuation 
depending on the energy. This makes their comparison with E866 data less
good than ours (we will obtain a similar result by fixing $M^2=M^2_{J/\Psi}$).
In a previous work \cite{kop} their comparison with E866 data where better
but they limited their analysis to $-0.1<x_F<0.25$ in order to see the influence
of the formation time.

To conclude, a high energy $pA$ collision
can be seen, in the laboratory frame, 
as the multiple scattering of the system of quarks and
gluons into which the incoming proton fluctuates. The coherence length of
this fluctuation increases with the energy and $x_F$ and two regions
can be distinguished. When $l_c\ll R_A$ no shadowing to gluons is present
and the suppression of $c\bar c$ pairs is given by the usual formula 
(\ref{eqnucabs}). When $l_c\gsim R_A$ the whole nucleus takes part in the
collision and typical coherent phenomena as shadowing appear. The main point
is that this change
of regime is accompanied by a change $\sigma^{abs}\to \widetilde\sigma$
in the expressions. If these two cross sections are different, the
suppression at large values of $x_F$ and/or larger energies is bigger.
Let's also comment about the $\Psi'$: experimental data from NA38 don't
see any difference between $J/\Psi$ and $\Psi'$ suppression, whereas some
difference seems to appear in E866/NuSea. 
This difference 
would be very easily accounted for in our approach just by taking the
absorptive cross sections different for singlet and octet $c\bar c$ states:
as the total $J/\Psi$ has a contribution of $\sim$40\% singlet coming
from disintegrations of $\chi_c$, a larger $\sigma^{abs}$ for octet
than for singlet would explain the difference. This would introduce 
a new parameter.
Finally, the extrapolations to RHIC and LHC energies depend on two assumptions,
the energy dependence of $\widetilde\sigma$ and $\sigma^{abs}$ (that we have 
taken as constants) and the value $M^2$ of the effective mass
of the fluctuation. With this last assumption we are wondering if
the $x_F$ scaling of the suppression will still be valid at high energy
(in the case $M^2/s\sim$ const.)
or if, on the contrary, a scaling in $x_2$ will appear, this last
possibility is the most reasonable as the proportionality in $s$ of
$M^2$ seems to be a finite energy effect.
Let's give some estimations for RHIC and LHC at $x_F\sim 0$. 
Assuming no energy dependence
for the cross sections $\widetilde\sigma$ and $\sigma^{abs}$ we obtain for
$pAu$ collisions at $\sqrt{s}=$200 GeV
a ratio over $pp$ of
0.81 if the scaling in $x_F$ is maintained and 0.43 if it is not.
This means ratios of 0.67 and 0.17 for $AuAu$ collisions respectively.
In the case of LHC, the only difference is shadowing, that will not affect
the case $M^2/s\sim$ const, in the other case, we obtain 0.31 for $pPb$
collisions and 0.1 for $PbPb$ at $\sqrt{s}=$5500 GeV. 

\noi {\bf Acknowledgments:}
I would like to thank N. Armesto, M. Braun, A. Capella, Yu. Dokshitzer,
A. Kaidalov,
C. Pajares, D. Schiff and X.-N. Wang for useful discussions, to
M. Leitch, C. Louren\c co and S. Rangod for information about experimental data 
and to
Ministerio de Educaci\'on y Cultura of Spain for financial support.

\newpage
\centerline{\bf \Large Figure captions:}
\vskip 1cm

\noi {\bf Fig. 1.}
Comparison of our results on $x_F$ dependence of nuclear suppression
(solid lines)
with E866 data \cite{e866}
for $Fe/Be$ and $W/Be$. Effects of shadowing as given by EKS98 parametrization
including absorption with 
$\sigma^{abs}$=
4.5 mb in eq. (\ref{eqnucabs}) are also shown (dashed lines).

\vskip 0.5cm
\noi {\bf Fig. 2.}
$B_{\mu\mu}\sigma^\psi$/A measured by NA38 Collaboration \cite{pAdata}
at $E_{lab}$=450 GeV$^2$ (black boxes) and $E_{lab}$=200 GeV$^2$ (white boxes)
compared with our results at $E_{lab}$=450 GeV$^2$ (solid line)
and $E_{lab}$=200 GeV$^2$ (dotted line). Also shown, the suppression given
by eq. (\ref{eqnucabs}) with $\sigma^{abs}$=3 mb.

\vskip 0.5cm
\noi {\bf Fig. 3.}
$\alpha$ parameter as a function of $x_F$ compared with data from
NA3 \cite{na3} at $E_{lab}$=200 GeV$^2$.

\newpage

\centerline{\bf Figure 1}
\vspace{1cm}

\begin{center}
\hspace{-1.2cm}\epsfig{file=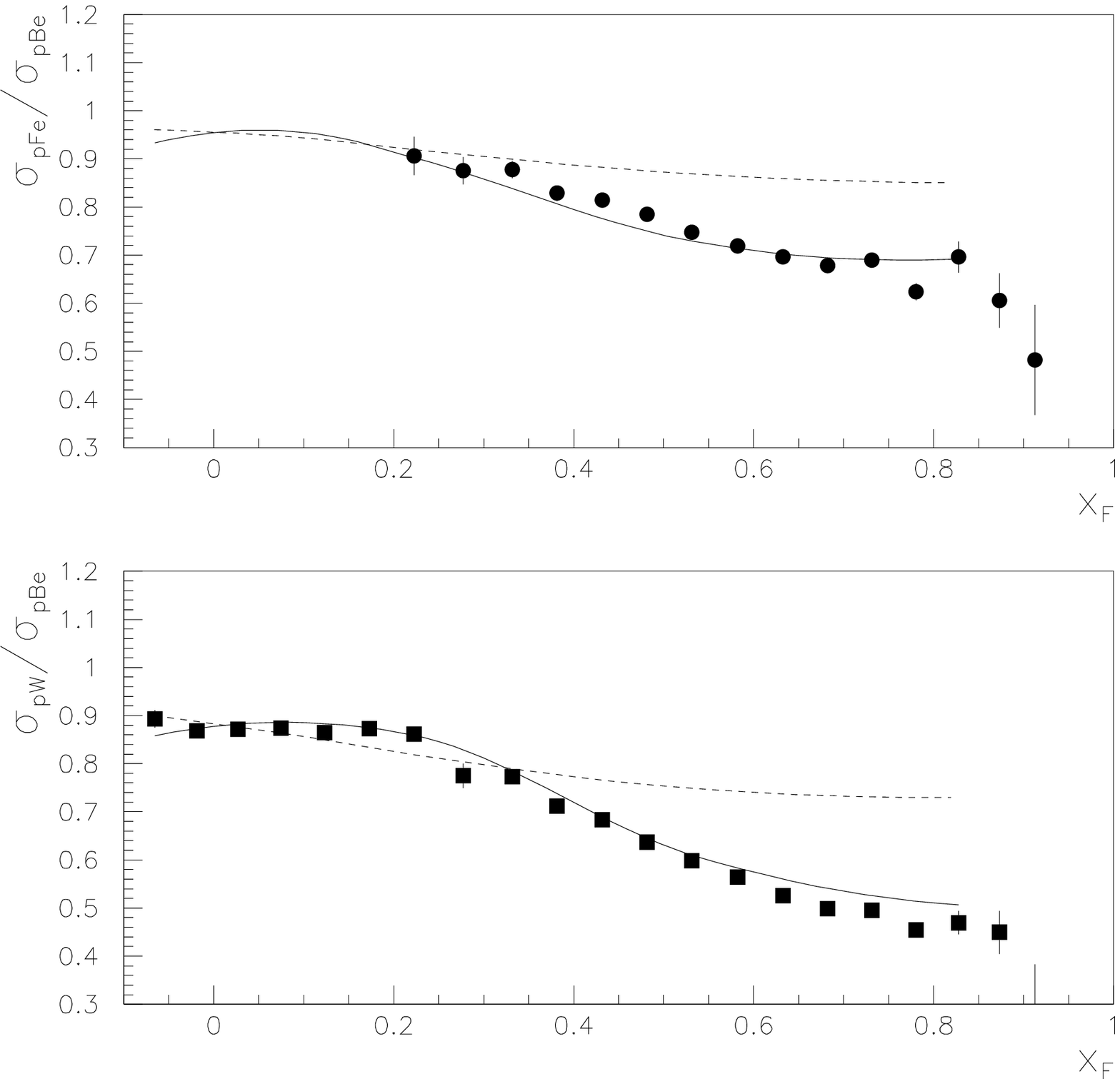,width=16.cm}
\end{center}
\newpage

\centerline{\bf Figure 2}
\vspace{1cm}

\begin{center}
\hspace{-1.2cm}\epsfig{file=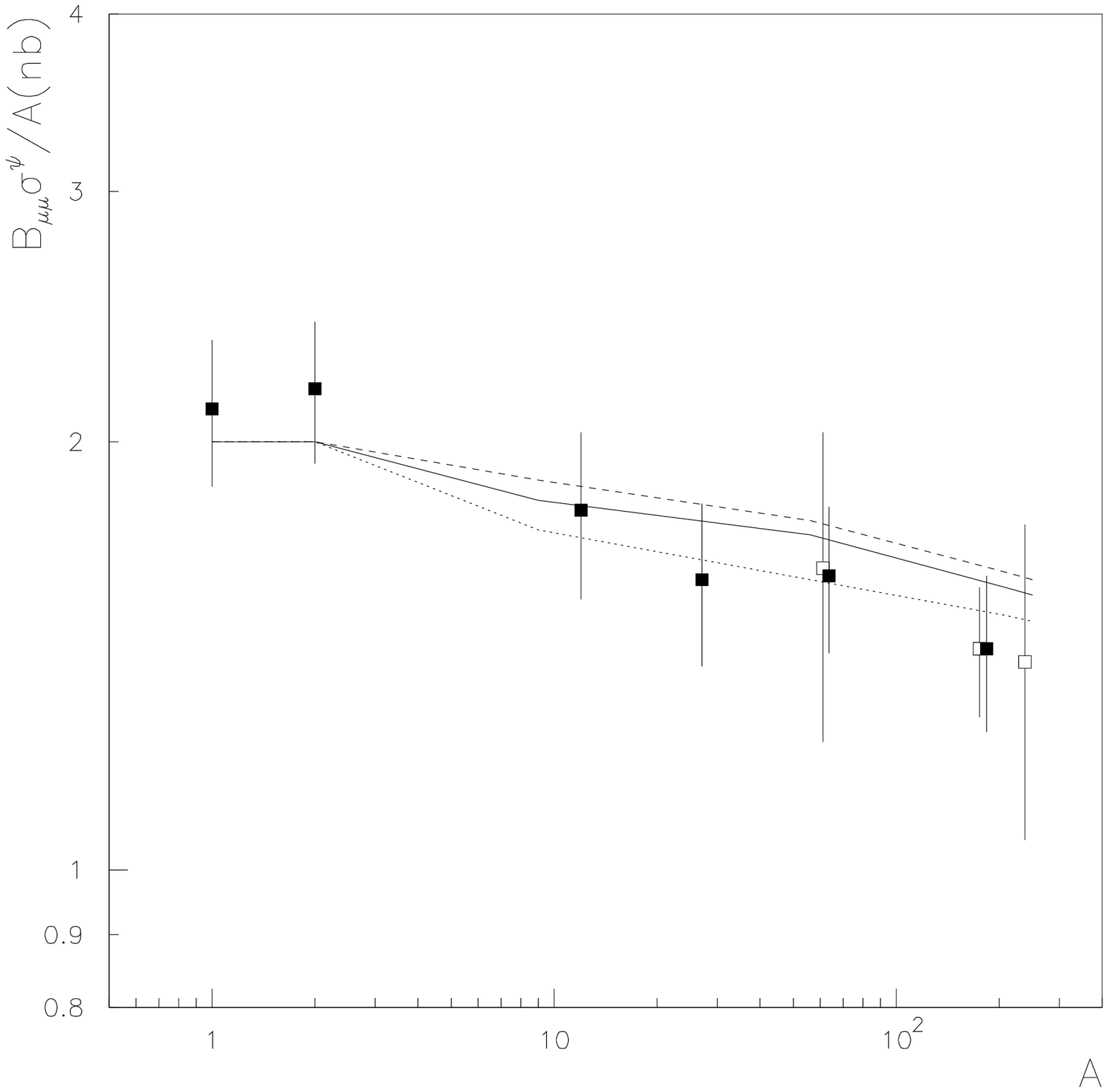,width=16.cm}
\end{center}
\newpage

\centerline{\bf Figure 3}
\vspace{1cm}

\begin{center}
\hspace{-1.2cm}\epsfig{file=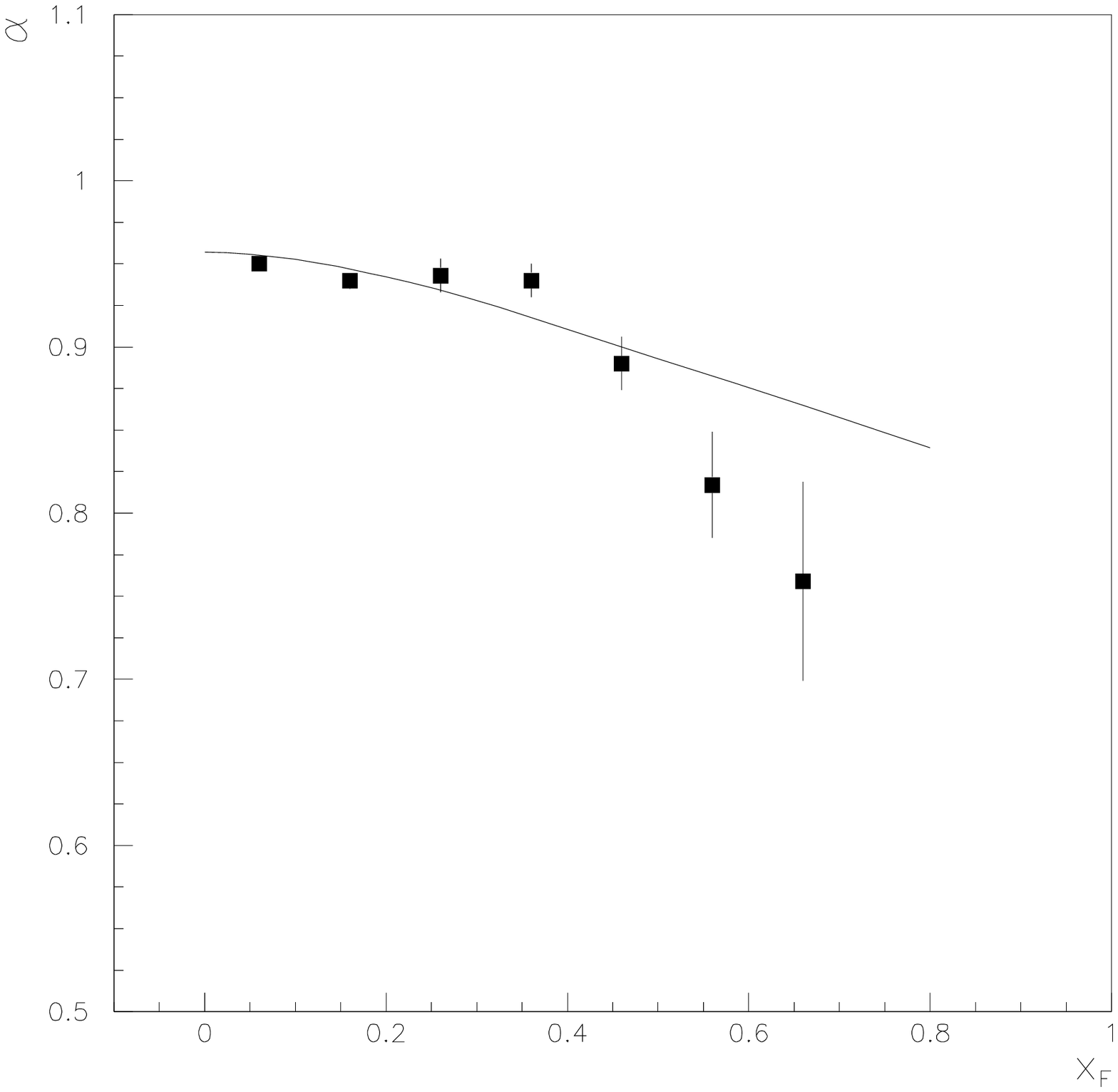,width=16.cm}
\end{center}


\newpage


\begin{thebibliography}{99}

\bibitem{matsuisatz}
T.~Matsui and H.~Satz,
Phys.\ Lett.\ B {\bf 178}, 416 (1986).

\bibitem{na50}
M.~C.~Abreu {\it et al.}  [NA50 Collaboration],
Phys.\ Lett.\ B {\bf 477}, 28 (2000).

\bibitem{deconf}
J.~Blaizot and J.~Ollitrault,
Phys.\ Rev.\ Lett.\ {\bf 77}, 1703 (1996);
J.~Blaizot, M.~Dinh and J.~Ollitrault,
Phys.\ Rev.\ Lett.\  {\bf 85}, 4012 (2000);
D.~Kharzeev, C.~Lourenco, M.~Nardi and H.~Satz,
Z.\ Phys.\ C {\bf 74}, 307 (1997);
M.~Nardi and H.~Satz,
Phys.\ Lett.\ B {\bf 442}, 14 (1998).

\bibitem{alfons}
N.~Armesto, A.~Capella and 
E.~G.~Ferreiro,
Phys.\ Rev.\ C {\bf 59}, 395 (1999);
A.~Capella, E.~G.~Ferreiro and A.~B.~Kaidalov,
Phys.\ Rev.\ Lett.\ {\bf 85}, 2080 (2000); 
A.~Capella, A.~B.~Kaidalov and D. Sousa,
nucl-th/0105021.

\bibitem{kharzeev}
D.~Kharzeev and H.~Satz,
Phys.\ Lett.\ B {\bf 366}, 316 (1996).

\bibitem{densities}
C.~W.~De Jager, H.~De Vries and C.~De Vries,
Atom.\ Data Nucl.\ Data Tabl.\  {\bf 14}, 479 (1974).

\bibitem{pAdata}
M.~C.~Abreu {\it et al.} [NA38 collaboration]
Phys.\ Lett.\ B {\bf 466}, 408 (1999).

\bibitem{e866}
M.~J.~Leitch {\it et al.}  [FNAL E866/NuSea collaboration],
Phys.\ Rev.\ Lett.\ {\bf 84}, 3256 (2000).

\bibitem{eks98}
K.~J.~Eskola, V.~J.~Kolhinen and C.~A.~Salgado,
Eur.\ Phys.\ J.\ C {\bf 9}, 61 (1999);
K.~J.~Eskola, V.~J.~Kolhinen and P.~V.~Ruuskanen,
Nucl.\ Phys.\ B {\bf 535}, 351 (1998).

\bibitem{mij}
N.~Armesto, M.~A.~Braun, A.~Capella, C.~Pajares and C.~A.~Salgado, 
Nucl.\ Phys.\ B {\bf 509}, 357 (1998).

\bibitem{perc}
N.~Armesto and C.~A.~Salgado,
hep-ph/0011352.

\bibitem{lowx}
A.~Capella, E.~G.~Ferreiro, A.~B.~Kaidalov and C.~A.~Salgado,
Phys.\ Rev.\ D {\bf 63}, 054010 (2001);
Nucl.\ Phys.\ B {\bf 593}, 336 (2001).

\bibitem{na3}
J.~Badier {\it et al.}  [NA3 Collaboration],
Z.\ Phys.\ C {\bf 20}, 101 (1983).

\bibitem{arleo}
F.~Arleo, P.~B.~Gossiaux, T.~Gousset and J.~Aichelin,
Phys.\ Rev.\ C {\bf 61}, 054906 (2000).

\bibitem{eloss}
S. Brodsky and P. Hoyer, Phys. Lett. B {\bf 298}, 165 (1993);
S. Gavin and J. Milana, Phys. Rev. Lett. {\bf 68}, 1834 (1992);
F.~Arleo, P.~B.~Gossiaux, T.~Gousset and J.~Aichelin, hep-ph/0005230;
M.~B.~Johnson {\it et al.}  [FNAL E772 Collaboration],
Phys.\ Rev.\ Lett.\  {\bf 86}, 4483 (2001); M.~B.~Johnson {\it et al.},
hep-ph/0105195.

\bibitem{domin}
R.~Baier, D.~Schiff and B.~G.~Zakharov,
Ann.\ Rev.\ Nucl.\ Part.\ Sci.\ {\bf 50}, 37 (2000).

\bibitem{hwa}
R.~C.~Hwa, J.~Pisut and N.~Pisutova,
Phys.\ Rev.\ Lett.\ {\bf 85}, 4008 (2000).

\bibitem{ramona}
R.~Vogt,
Phys.\ Rev.\ C {\bf 61}, 035203 (2000); 
R.~Vogt, S.~J.~Brodsky and P.~Hoyer,
Nucl.\ Phys.\ B {\bf 360}, 67 (1991). 

\bibitem{kaidalov}
K.~Boreskov, A.~Capella, A.~Kaidalov and J.~Tran Thanh Van,
Phys.\ Rev.\ D {\bf 47}, 919 (1993).

\bibitem{kopnew}
B.~Z.~Kopeliovich, A. Tarasov and J.~Hufner,
hep-ph/0104256.

\bibitem{kop}
Y.~B.~He, J.~Hufner and B.~Z.~Kopeliovich,
Phys.\ Lett.\ B {\bf 477}, 93 (2000).


\end{thebibliography}
\end{document}